\documentclass{cupconf}
\usepackage{epsfig}
\usepackage{subfigure}

  \checkfont{eurm10}
  \iffontfound
    \IfFileExists{upmath.sty}
      {\typeout{^^JFound AMS Euler Roman fonts on the system,
                   using the 'upmath' package.^^J}%
       \usepackage{upmath}}
      {\typeout{^^JFound AMS Euler Roman fonts on the system, but you
                   dont seem to have the}%
       \typeout{'upmath' package installed. cupconf.cls can take advantage
                 of these fonts,^^Jif you use 'upmath' package.^^J}%
      }
  \else
  \fi

  \checkfont{msam10}
  \iffontfound
    \IfFileExists{amssymb.sty}
      {\typeout{^^JFound AMS Symbol fonts on the system, using the
                'amssymb' package.^^J}%
       \usepackage{amssymb}%
         \let\leq=\leqslant
         
      }{}
  \fi

  \IfFileExists{amsbsy.sty}
    {\typeout{^^JFound the 'amsbsy' package on the system, using it.^^J}%
     \usepackage{amsbsy}}
    {}

\newsavebox{\astrutbox}
\sbox{\astrutbox}{\rule[-5pt]{0pt}{20pt}}

\newcommand\etal{\mbox{\textit{et al.}}}

\def\etal{{et al.\ }}
\def\gsim{ \lower .75ex \hbox{$\sim$} \llap{\raise .27ex \hbox{$>$}} }
\def\lsim{ \lower .75ex\hbox{$\sim$} \llap{\raise .27ex \hbox{$<$}} }
\def\msun{\,{\rm M_\odot}}

\def\simlt{\mathrel{\rlap{\lower 3pt\hbox{$\sim$}}\raise 2.0pt\hbox{$<$}}}
\def\simgt{\mathrel{\rlap{\lower 3pt\hbox{$\sim$}} \raise 2.0pt\hbox{$>$}}}
\def\gtrsim{\lower.5ex\hbox{\gsim}}
\def\lesssim{\lower.5ex\hbox{\lsim}}

\def\kms{{\rm\,km\,s^{-1}}}

\def\gcm3{{\rm g\,\, cm^{-3}}}
\def\mbulge{M_{\rm Bulge}}
\def\msunpc3{\msun~{\rm {pc^{-3}}}}

\newcommand{\be}{\begin{equation}}
\newcommand{\ee}{\end{equation}}

\title[Binary Black Holes]
{Birth of massive black hole binaries}

\author[Monica Colpi et al.]
{%
M\ls O\ls N\ls I\ls C\ls A\ls  \ns C\ls O\ls L\ls P\ls I$^1$, \ns
M\ls A\ls  S\ls S\ls I\ls M\ls O \ns D\ls O\ls T\ls T\ls I$^2$, 
\ns L\ls U\ls C\ls I\ls O\ls \ns M\ls A\ls Y\ls E\ls R$^3$ \ns \and\ns
S\ls  T\ls  E\ls  L\ls  I\ls  O\ls  S\ls \ns  K\ls  A\ls  Z\ls  A\ls  
N\ls  T\ls  Z\ls  I\ls  D\ls  I\ls  S$^4$\ls
} 

\affiliation{ $^{(1)}$Department of Physics G. Occhialini, University
  of Milano Bicocca, Milano, Italy, $^{(2)}$Department of Physics,
  University of Insubria, Como, Italy, $^{(3)}$Insitute of Theoretical
  Physics, Zurich, Switzerland, $^{(4)}$ Kavli Institute for Particle
  Astrophysics and Cosmology, Department of Physics, Stanford
  University, USA}

\pubyear{2007}
\volume{000}
\pagerange{000--000}
\date{?? and in revised form ??}
\begin{document}

\maketitle

\begin{abstract}
  If massive black holes (BHs) are ubiquitous in galaxies and galaxies
  experience multiple mergers during their cosmic assembly, then BH
  binaries should be common albeit temporary features of most galactic
  bulges. Observationally, the paucity of active BH pairs points
  toward binary lifetimes far shorter than the Hubble time, indicating
  rapid inspiral of the BHs down to the domain where gravitational
  waves lead to their coalescence.  Here, we review a series of
  studies on the dynamics of massive BHs in gas-rich galaxy mergers
  that underscore the vital role played by a cool, gaseous component
  in promoting the {\it rapid formation of the BH binary}.  The BH
  binary is found to reside at the center of a massive
  self-gravitating nuclear disc resulting from the collision of the
  two gaseous discs present in the mother galaxies.  Hardening by
  gravitational torques against gas in this grand disc is found to
  continue down to sub-parsec scales.  The eccentricity decreases with
  time to zero and when the binary is circular, accretion sets in
  around the two BHs.  When this occurs, each BH is endowed with it
  own small-size ($\simlt 0.01$ pc) accretion disc comprising a few
  percent of the BH mass. Double AGN activity is expected to occur on
  an estimated timescale of $\simlt 1$ Myr.  The double nuclear
  point--like sources that may appear have typical separation of
  $\simlt 10$ pc, and are likely to be embedded in the still ongoing
  starburst.  We note that a potential threat of binary stalling, in a
  gaseous environment, may come from radiation and/or mechanical
  energy injections by the BHs.  Only short--lived or sub--Eddington
  accretion episodes can guarantee the persistence of a dense cool gas
  structure around the binary necessary for continuing BH inspiral.

\end{abstract}

\firstsection 

\section{Introduction}

Dormant black holes (BHs) with masses in excess of $\simgt 10^6\msun$
are ubiquitous in bright galaxies today (Kormendy \& Richstone 1995;
Richstone 1998).  Relic of an earlier active phase as
quasars, these massive BHs appear a clear manifestation of the cosmic
assembly of galaxies.  The striking correlations observed between the
BH masses and properties of the underlying hosts
(Magorrian \etal 1998; Ferrarese \& Merritt 2000; Gebhardt et al. 2000; Graham
\& Driver 2007) indicates unambiguously that BHs evolve in symbiosis
with galaxies, affecting the environment on large--scales 
and self-regulating their growth (Silk \& Rees 1998; King 2000; 
Granato et al. 2004; Di Matteo, Springel \& Hernquist, 2005).

According to the current paradigm of structure formation, galaxies
often interact and collide as their dark matter halos assemble in a
hierarchical fashion (Springel, Frenk \& White 2006), and BHs
incorporated through mergers into larger and larger systems are
expected to evolve concordantly (Volonteri, Haardt \& Madau 2003).
In this astrophysical context, close BH {\it pairs} form as natural
outcome of binary galaxy mergers (Kazantzidis et al. 2005).

In our local universe, one outstanding example is the case of the
ultra--luminous infrared galaxy NGC 6240, an ongoing merger between two
gas--rich galaxies (Komossa \etal 2003; for a review on binary black
holes see also Komossa 2006). {\it Chandra}
images have revealed the occurrence of two nuclear X-ray sources, 1.4
kpc apart, whose spectral distribution is consistent with being two
accreting massive BHs embedded in the diluted emission of a starburst.  
Similarly, Arp 299 (Della Ceca et al. 2002; Ballo et
al. 2004) is an interacting system hosting an
obscured active nucleus, and possibly a second less
luminous one, distant several kpc away.  A third example is 
the elliptical galaxy 0402+369 where the cleanest case of a massive
BH {\it binary} has been recently discovered. Two compact variable,
flat--spectrum active nuclei are seen at a projected separation of
only 7.3 pc (Rodriguez et al. 2006). Arp 299, NGC 6240, and 0402+369
may just highlight different stages of the BH dynamical evolution
along the course of a merger, with 0402+369 being the latest, most
evolved phase (possibly related to a dry merger).  Energy and angular
momentum losses due to gravitational waves are not yet significant in
0402+392, so that stellar interactions and/or material and
gravitational torques are still necessary to bring the BHs down to the
domain controlled by General Relativity.

From the above considerations and observational findings, it is clear
that binary BH inspiral down to coalescence is a major astrophysical
process that can occur in galaxies.  It is accompanied by a
gravitational wave burst so powerful to be detectable out to very
large redshifts with current planned experiments like the Laser
Interferometer Space Antenna ({\it LISA}; Bender et al. 1994; Vitale
et al. 2002).  These extraordinary events will provide not only a firm
test of General Relativity but also a view, albeit indirect, of galaxy
clustering (Haehnelt 1994; Jaffe \& Backer 2003; Sesana et al. 2005).
With {\it LISA}, BH masses and spins will be measured with such an
accuracy (Vecchio 2004) that it will be possible to trace the BH mass
growth across all epochs. Interestingly, {\it LISA} will explore a
mass range between $10^3\msun$ and $10^7\msun$ that is complementary
to that probed by the distant massive quasars ($>10^7\msun$),
providing a complete census of the BHs in the universe.

Both minor as well as major mergers with BHs accompany galaxy
evolution in environments that involve either gas-rich (wet) as well
as gas-poor (dry) galaxies.  Thus, the dynamical response of galaxies
to BH pairing should differ in many ways according to their
properties.  Exploring the expected diversities in a self-consistent
cosmological scenario is a major challenge and only recently, with the
help of high-resolution N-body/SPH simulations, it has become possible
to ``start'' addressing a number of compelling issues.  Galaxy mergers
cover cosmological volumes (a few to hundred kpc aside), whereas BH
mergers probe volumes of only few astronomical units or less. Thus, tracing
the BH dynamics with scrutiny requires N-Body/SPH force resolution
simulations spanning more than nine orders of magnitude in length.
For this reason, two complementary approaches have been followed in the 
literature.  A statistical approach (based either on Monte Carlo
realizations of merger trees or on N-Body/SPH large scale simulations)
follows the collective growth of BHs inside dark matter
halos.  Supplemented by semi-analytical modeling of BH dynamics
(Volonteri et al. 2003) or/and by sub-grid resolution criteria for
accretion and feedback (Springel \& Hernquist 2003; Springel, Di
Matteo \& Hernquist 2005), these studies have proved to be powerful
in providing estimates of the expected coalescence rates, and in
tracing the overall cosmic evolution of BHs including their feedback
on the galactic environment (Di Matteo, Springel \& Hernquist 2005; Di Matteo
\etal 2007).  The second approach, that we have been following, looks
at individual binary collisions, as it aims at exploiting in detail
the BH dynamics and some bulk physics from the galactic scale down to
and within the BH sphere of influence.  Both approaches, the
collective and the individual, are necessary and complementary, the
main challenge being the implementation of realistic input physics in
the dynamically active environment of a merger.

Following a merger, how can BHs reach the gravitational wave inspiral
regime?  The overall scenario was first outlined by Begelman,
Blandford \& Rees (1980) in their seminal study on the dynamical
evolution of BH pairs in pure stellar systems.  They indicated three
main roots for the loss of orbital energy and angular momentum: (I)
dynamical friction against the stellar background acting on each
individual BH; (II) hardening via 3--body scatterings off single stars
when the BH binary forms; (III) gravitational wave back--reaction.

Early studies explored phase (I) simulating the collisionless merger
of spherical halos (Makino \& Ebisuzaki 1996; Milosavljevi\'c \&
Merritt 2001; Makino \& Funato 2004).  Governato, Colpi \& Maraschi
(1994) in particular first noticed that when two equal mass halos
merge, the twin BHs nested inside the nuclei are dragged
effectively toward the center of the remnant galaxy by dynamical
friction and form a close pair, but that the situation reverses in
unequal mass mergers, where the less massive halo tidally disrupted
leaves its ``naked'' BH wandering in the outskirts of the main halo.
Thus, depending on the halo mass ratio and internal structure, the
transition from phase (I) to phase (II) can be prematurely aborted or
drastically relented.  Similarly, the transit from phase (II) to phase
(III) is not always secured, as the stellar content inside the ``loss
cone" may not be rapidly refilled with fresh low--angular momentum
stars to harden the binary down to separations where gravitational
wave driven inspiral sets in (see, e.g., Milosavljevic \& Merritt
2001; Yu 2002; Berczik, Merritt \& Spurzem 2005; Sesana, Haardt \&
Madau 2007). For an updated review on the last parsec problem and its
possible solution (see Merritt 2006a; Gualandris \& Merritt 2007).

Since BH coalescences are likely to be events associated with
mergers of (pre--)galactic structures at high redshifts, it is likely that
their dynamics occurred in gas dominated backgrounds, NGC 6240 being
just the most outstanding case visible in our local universe. 
Other processes of BH binary hardening are expected to operate in 
presence of a dissipative gaseous component that we will highlight and study
here.   

Kazantzidis \etal (2005) first explored the effect of gaseous
dissipation in mergers between gas--rich disc galaxies with central
BHs, using high resolution N--Body/SPH simulations.  They found that
the merger triggers large--scale gas dynamical instabilities that lead
to the gathering of cool gas deep in the potential well of the
interacting galaxies.  In minor mergers, this fact is essential in
order to bring the BHs to closer and closer distances before the less
massive galaxy, tidally disrupted, is incorporated in the main galaxy.
Moreover, the interplay between strong gas inflows and star formation
leads naturally to the formation, around the two BHs, of a grand,
massive ($\sim 10^9\msun$) gaseous disc on a scale smaller than $\sim
100$ pc.  It is in this equilibrium circum--nuclear disc that the
dynamical evolution of the BHs continues, after the merger has been
subsided.  Escala \etal (2005, hereinafter ELCM05; see also Escala
\etal 2004) have been the first to study the role played by gas in
affecting the dynamics of massive ($\sim 10^8\msun$) twin BHs in
equilibrium Mestel discs of varying clumpiness.  In both these
approaches (i.e., in the large scale simulations of Kazantzidis et
al., and in the equilibrium disc models of ELCM05) it was clear that
the gas temperature is a key physical parameter and that 
a hot gas brakes the BHs inefficiently. Instead, when the gas is
allowed to cool, the drag becomes efficient: the large enhancement of
the local gas density relative to the stellar one leads to the
formation of prominent density wakes that are decelerating the BHs
down to the scale where they form a ``close'' binary.  Later, binary
hardening occurs under mechanisms that are only partially explored,
and that are now subject of intense investigation.  The presence of a
cool circum--binary disc and of small--scale discs around each
individual BH appear to be critical for their evolution down to the
domain of gravitational waves driven inspiral. In this context there
is no clear ``stalling problem'' that emerges from current
hydrodynamical simulations but this critical phase need a more
through, coherent analysis.

The works by Kazantzidis \etal (2005) and ELCM05 
have provided our main motivation to study
the process of BH pairing along two lines:
In gas-rich binary mergers, line (1) aims at studying 
the transit from state (A) of pairing when
each BH moves individually inside the time-varying
potential of the colliding galaxies, to state (B) when the two BHs
dynamically couple their motion to form a binary.
The transit from (A) $\to$ (B) requires exploring a dynamic
range of five orders of magnitude in length from the cosmic
scale of a galaxy merger of 100 kpc down to the parsec scale
for BHs of million solar masses (i.e., BHs in the LISA
sensitivity domain).  After all
transient inflows have subsided and a
new galaxy has formed, the BH binary is expected to enter phase (C) 
where it hardens under the action of gas-dynamical and gravitational torques.
Research line (2) aims at studying the braking of the BH binary
from (B) $\to$ (C) and further in, exploring
the possibility that during phase (C) two discs form  
and grow around each individual 
BH. 
As first discussed by Gould \& Rix (2000) 
the binary may later enter a new phase (D) 
controlled by the balance of viscous and gravitational torques
in a circum--binary disc surrounding the BHs, in a manner analogous to the
migration of planets in circum-stellar discs (a scenario particularly
appealing when the BH mass ratio is less than unity). 
Phase (D) likely evolves into (E) when gravitational wave inspiral
terminates the BH binary evolution.

There is a number of key questions to address:

\noindent
(i) How does transition from state (A) $\to$ (B) depend on the gas 
thermodynamics? How do BHs bind?

\noindent
(ii) In the grand nuclear disc inside the remnant galaxy, how do
eccentric orbits evolve? Do they become circular or highly
eccentric?

\noindent
(iii) During the hardening through phase (B) and (C), do the BHs   
collect substantial amounts of gas to form cool individual discs?  

\noindent
(iv) Can viscous torques drive the binary into
the gravitational wave decaying phase?

\noindent
(v) Is there a threat of a {\it stalling} problem when transiting
from (C) $\to$ (D) or from (D) $\to$ (E)?  And, for which mass ratios
and ambient conditions?

\section{Dynamics of BHs in disc--galaxy mergers}

In this section, we track the large--scale dynamics of two massive BHs
during the merger between two gas--rich (equal mass) disc galaxies,
and later focus on the process leading to the formation of a Keplerian
BH binary.

\subsection {Modeling galaxy mergers}

We start simulating, with the N-Body/SPH code {\sl Gasoline} (Wadsley,
Stadel \& Quinn 2004), the collision between two galaxies, similar to
the Milky Way, comprising a stellar bulge, a disc of stars and gas,
and a massive, extended spherical dark matter halo with NFW density
profile (Navarro, Frenk \& White 1996; Klypin, Zhao \& Somerville
2002).  The halo has a virial mass $M_{\rm vir}=10^{12}\msun,$
concentration parameter $c=12$ and dimensionless spin parameter
$\lambda=0.031$ consistent with current structure formation models.
The disc of mass $M_{\rm disc}=0.04M_{\rm vir}$ has a surface density
distribution that follows an exponential law with scale length of 3.5
kpc and scale height 350 pc. The spherical bulge (Hernquist 1993)
has mass $M_{\rm bulge}=0.008 M_{\rm vir}$ and scale radius of
700 pc.  Initially, the dark matter halo has been adiabatically
contracted to respond to the growth of the disc and bulge, resulting in
a model with a central density slope close to isothermal.
Each galaxy consists of $10^5$ stellar disc particles, $10^5$ bulge
particles, and $10^6$ halo particles.  The gas fraction, $f_{\rm g}$,
is 10\% of the total disc mass and is represented by $10^5$ particles
($10^6$ in a refined simulation).  To each of the galaxy model we
added a softened particle, initially at rest,
to represent the massive BH  at the center of the bulge.  
The BH mass is $M_{\rm BH}=2.6 \times
10^{6} M_{\odot}$, according to $M_{\rm BH}$--$\sigma$ relation.  In
the major merger, the BHs are twin BHs (i.e., the mass ratio $q_{\rm
  BH}=1$).

Different encounter geometries were explored in the simulations by
Kazantzidis et al. (2005): they comprise prograde or retrograde
coplanar mergers as well as mergers with galactic discs inclined
relative to the orbital plane.  The simulation presented in this
proceeding refers to a coplanar prograde encounter. This particular
choice is by no means special for our purpose, except that the
galaxies merge slightly faster than in the other cases thus limiting
our computational burden.  The galaxies approach each other on
parabolic orbits with pericentric distances equal to 20\% of the
galaxy's virial radius, typical of cosmological mergers (Khochfar \&
Burkert 2006).  The initial separation of the halo centers is twice
their virial radii and their initial relative velocity is determined
from the corresponding Keplerian point--mass orbit.

We include radiative cooling from a primordial mixture of hydrogen
and helium, and adopt the star formation algorithm by Katz (1992) where
gas particles, in dense cold Jeans unstable regions and in convergent
flows, spawn N-body particles at a rate proportional to the local
dynamical time (with star formation efficiency of 0.1).  Radiative
cooling is switch off at a relatively high floor temperature of
20,000 K to account for turbulent heating, non--thermal
pressure forces, and the presence of a warm interstellar medium. In
this large--scale simulation, the force resolution is $\sim 100$
pc.

The computational volume is later refined during the final stage of
the merger with the technique of static particle splitting (Kaufmann
\etal 2006) in order to achieve a resolution of $2$ pc.  The
fine-grained region is large enough to guarantee that the dynamical
timescale of the entire coarse-grained region is much longer than the
dynamical timescale of the refined region so that gas particles from
the coarse region will reach the fine region on a timescale longer
than the actual time span probed in this work.  In the refined
simulations stars and dark matter particles essentially provide a
smooth background potential, while the computation focuses on the gas
component which dominates by mass in the nuclear region.
A volume of 30 kpc in radius is selected  while the two galaxy cores are
separated by only $6$ kpc. 
Inside this region, the simulation is
carried on with as many as $2\times 10^6$ gas particles.
The mass resolution in the gas component, 
originally of $2 \times 10^4 M_{\odot},$ now becomes $\sim 3000 M_{\odot}$
after splitting.

A starburst with a peak star formation rate of $\sim 30\msun$ yr$^{-1}$ 
takes
place when the cores finally merge and it is in this environment that
the BHs couple to form a binary.  The short dynamical timescale
involved in this process compared to the starburst duration ($\sim
10^8$ yr) suggest to model the thermodynamics and radiation physics
simply via an effective equation of state.  Calculations that include
radiative transfer show that the thermodynamic state of a metal rich
gas heated by a starburst can be approximated by an equation of state
of the form $P=(\gamma - 1) \rho u\,$ with $\gamma=7/5$ (Spaans \&
Silk 2000).  The specific internal energy $u$ evolves with time as a
result of $PdV$ work and shock heating modeled via the standard
Monaghan artificial viscosity term.  Shocks are generated even when a
self-gravitating disc forms with strong spiral arms.  The highly
dynamical regime modeled here is different from that considered in the
next section which could be evolved using an adiabatic equation of
state.  With this prescription we treat the gas as a one-phase medium
whose mean density and internal energy (the sum of thermal and
turbulent energy) correspond to the mean density and line width seen
in observed nuclear discs (Downes \& Solomon 1998).

\begin{figure}
\begin{center}
\includegraphics[height=0.50\textheight]{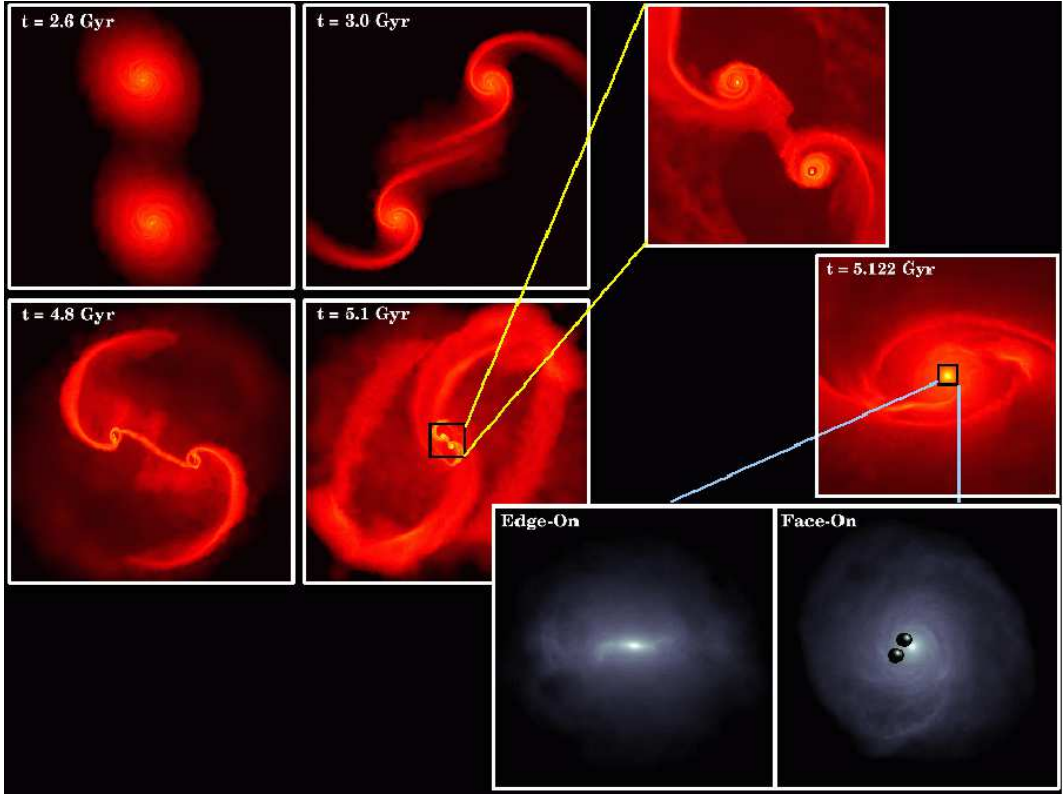}
\end{center}
\caption[]{
  The different stages of the merger between two identical disc
  galaxies seen face--on. The color coded density maps of the gas
  component are shown using a logarithmic scale, with brighter colors
  for higher densities.  The four panels to the left show the
  large-scale evolution at different times. The boxes are 120 kpc on a
  side (top) and 60 kpc on a side (bottom) and the density ranges
  between $10^{-2}$ atoms cm$^{-3}$ and $10^{2}$ atoms cm$^{-3}$.
  During the interaction tidal forces tear the galactic discs apart,
  generating spectacular tidal tails and plumes.  The panels to the
  right show a zoom in of the very last stage of the merger, about 100
  million years before the two cores have fully coalesced (upper
  panel), and 2 million years after the merger (middle panel), when a
  massive, rotating nuclear gaseous disc embedded in a series of
  large-scale ring-like structures has formed.  The boxes are now 8
  kpc on a side and the density ranges between $10^{-2}$ atoms
  cm$^{-3}$ and $10^{5}$ atoms cm$^{ -3}$.  The two bottom panels,
  with a gray color scale, show the detail of the inner 160 pc of
  the middle panel; the nuclear disc is shown edge-on (left) and
  face-on (right), and the two BHs are also shown in the face-on
  image.}
\end{figure}

\subsection {Large--scale dynamics and BH pairing}

The galaxies first experience few close fly-bys before merging.  In
these early phase of the collision, the cuspy potentials of both
galaxies are deep enough to allow for the survival of the baryonic
cores where the BHs reside. As the two dark matter halos sink into one
another under the action of dynamical friction, strong spiral patterns
appear in both the stellar and the gaseous discs. Non--axisymmetric
torques redistribute angular momentum, and as much as 60\% of the gas
originally present in the discs of the parent galaxies is funneled
inside the inner few hundred parsecs of each core.  This is
illustrated in the upper right panel of Figure 1, where the enlarged
color coded density map of the gas is shown, after 5.1 Gyr from the
onset of the collision. {\it Each of the two BHs are found
to be  surrounded by a
rotating stellar and gaseous disc of mass $\sim 4 \times 10^8$
M$_{\odot}$ and size of a few hundred parsecs.}  The two discs and BHs
are just 6 kpc far apart. Meantime a starburst of $\sim 30 \msun\,$yr$^{-1}$
has invested the
central region of the merger.

It is tempting to imagine that an episode of accretion onto each BH
starts at this time, similar to that observed in NGC 6240 (see Colpi
et al. 2007, for a discussion on accretion excited along the course of
a large--scale merger).  The {\it double} AGN activity in NGC 6240
occurs just on a similar scale, as the X--ray nuclei are 1 kpc apart
in projection on the sky.  It is remarkable that high--resolution
near--infrared images at the Keck II telescope, combined with radio
and X-ray positions, have revealed the habitat of the two active BHs
in this ultra--luminous system.  Each active BH appears to be at the
center of a rotating stellar disc surrounded by a cloud of young star
clusters lying in the plane of each disc (Max, Canalizo \& de Vries
2007). This hints to a consumption of a fraction of the gas disc into
stars along the course of the major merger and to BH fuelling by the
winds of these young stars.

As the interaction proceeds in the simulation, the two baryonic discs
around each BH get closer and closer, and eventually merge in a single
structure: a massive circum--nuclear disc. This is illustrated in Figure
1, in the mid right panel. The two BHs are now at a relative
separation comparable to the softening length of $\sim 100$ pc.  At
this stage we stop the simulation and start the one with increased
resolution.

\subsection {Formation of a circum--nuclear disc}

\begin{figure}
\begin{center}
\includegraphics[height=0.50\textheight]{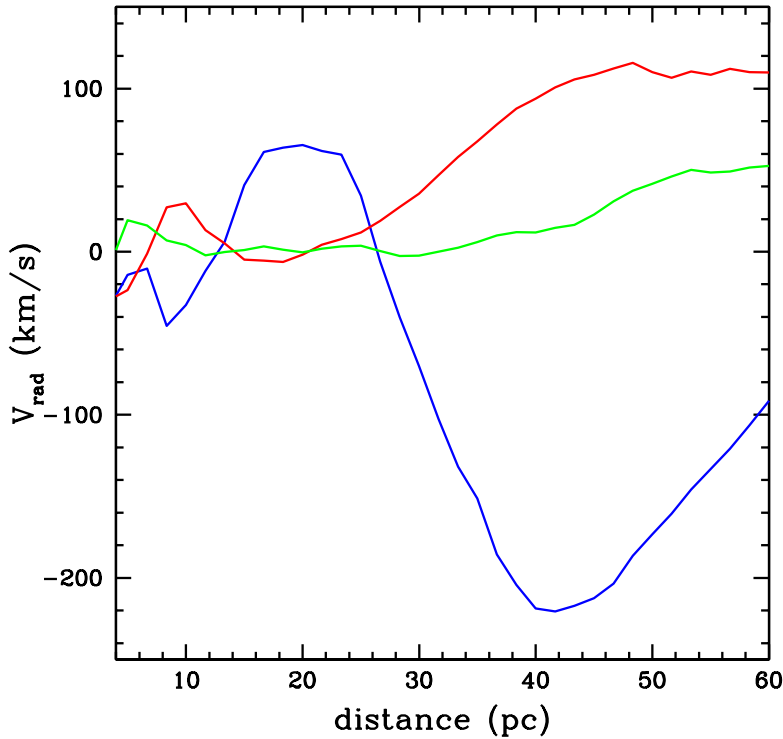}
\end{center}
\caption[]{
  Radial velocities within the nuclear disc ($\gamma = 7/5$) starting
  at $t=5.1218$ Gyr (blue line), and then after another $10^5$ yr 
  (red line) and $2 \times 10^5$ yr (green line). Remarkable inflow
  and outflow regions are the result of streaming motions within the
  bar and spiral arms arising in the disc during the phase of
  non-axisymmetric instability sustained by its self-gravity.  At
  later times the instability saturates due to self-regulation, and the
  radial motions also level down (green line).  }
\label{fig:birth}
\end{figure}

The gaseous cores merge in a single nuclear disc with mass of $3\times
10^9 M_{\odot}$ and size of $\simlt 100$ pc.  This {\it grand disc} is
more massive than the sum of the two nuclear cores formed earlier
because radial gas inflows occur in the last stage of the galaxy
collision.  A strong spiral pattern in the disc produces remarkable
radial velocities whose amplitude declines as the spiral arms weaken
over time.  Just after the merger, when non-axisymmetry is strongest,
radial motions reach amplitudes of $\sim 100\, \kms$ (see Figure 2).
This phase lasts only a couple of orbital times, while later the disc
becomes smoother as spiral shocks increase the internal energy which
in turn weakens the spiral pattern.  Inward radial velocities of order
$30-50\,\kms$ are seen for the remaining few orbital times.  .

The disc is surrounded by several rings and by a more diffuse,
rotationally-supported envelope extending out to more than a $\sim$
kpc from the center (Figure 1). A background of dark matter and stars
distributed in a spheroid is also present but the gas component is
dominant in mass within $\sim 300$ pc from the center.

The grand disc is rotationally supported ($v_{\rm rot} \sim 300$ km
s$^{-1}$) and also highly turbulent, having a typical velocity $v_{\rm
  turb} \sim 100$ km s$^{-1}$.  Multiple shocks generated as the cores
merge are the main source of this turbulence.  The disc is composed by
a very dense, compact region of size about 25 pc which contains half
of its mass (the mean density inside this region is $> 10^5$ atoms
cm$^{-3}$). The outer region instead, from 25 to 75-80 pc, has a
density 10-100 times lower, and is surrounded by even lower--density
rotating rings that extend out to a few hundred parsecs.  The disc
scale height also increases from inside out, ranging from 20 pc to
nearly 40 pc.  The volume-weighted density within 100 pc is in the
range $10^3-10^4$ atoms cm$^{-3}$, comparable to that of observed
nuclear disc (Downes \& Solomon 1998).  This suggests that the degree
of dissipation implied by our equation of state is a reasonable
assumption despite the simplicity of the thermodynamical scheme
adopted.

\subsection {Birth of a  BH binary}

The BHs have been dragged together with their cores toward the
dynamical center of the merging galaxies under the action of dynamical
friction, and now move inside the grand disc.

\begin{figure}
\begin{center}
\includegraphics[height=0.50\textheight]{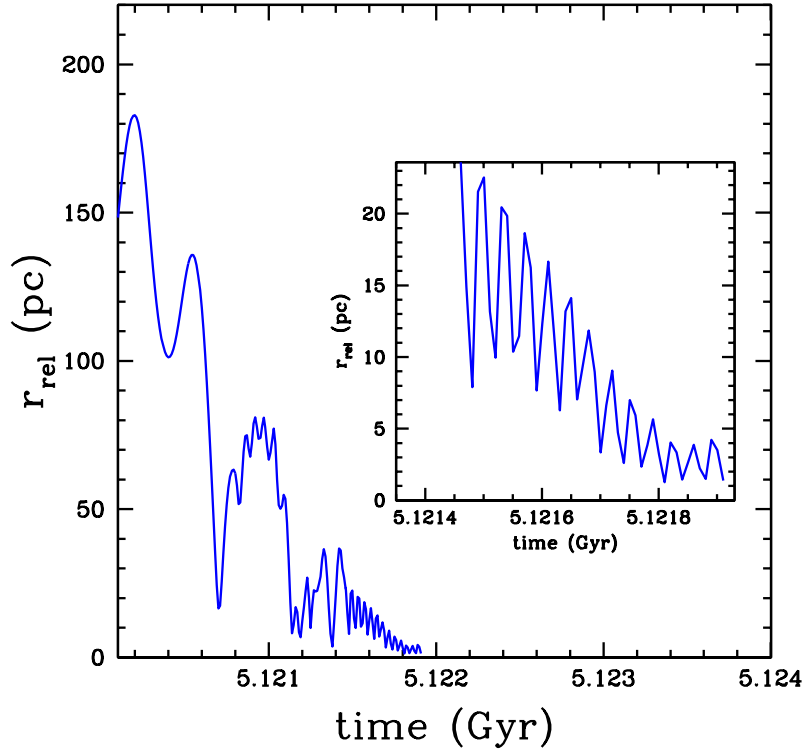}
\end{center}
\caption[]{Orbital separation of the two BHs as a function 
  of time during the last stage of the galaxy merger shown in Figure
  1.  The orbit of the pair is eccentric until the end of the
  simulation. The two peaks at scales of tens of parsecs at around
  $t=5.1213$ Gyr mark the end of the phase during which the two holes
  are still embedded in two distinct gaseous cores. Until this point
  the orbit is the result of the relative motion of the cores combined
  with the relative motion of each BH relative to the surrounding
  core, explaining the presence of more than one orbital frequency.
  The inset shows the details of the last part of the orbital
  evolution, which takes place in the nuclear disc arising from the
  merger of the two cores. The binary stops shrinking when the
  separation approaches the softening length (2 pc).}
\label{fig:birth}
\end{figure}

They keep sinking down from about 40 pc to a few pc, our
resolution limit.  We find that
{\it in less than a million years after the merger, 
the two holes are gravitationally bound to each other, as the
mass of the gas enclosed within their separation is less than the mass
of the binary. It is the gas that controls the orbital decay, not the stars.} 
Dynamical friction against the stellar background would bring the two
BHs this close only on a much longer timescale, $\sim 5 \times
10^7$ yr (Mayer et al. 2007, supporting online material). This short 
sinking timescale 
comes from 
the combination of (1) the fact that gas densities are much higher
than stellar densities in the center, and (2) that in the
mildly supersonic regime the drag against a gaseous background is stronger than
that in a stellar background with the same density (Ostriker 1999).  Adding star
formation is unlikely to change this conclusion as in our 
low-resolution galaxy merger simulations, the starburst timescale
of $\sim  10^8$ yr is much longer than the
binary formation timescale.

\subsection{Effect of thermodynamics of the
BH sinking}

\begin{figure}
\begin{center}
\includegraphics[height=0.50\textheight]{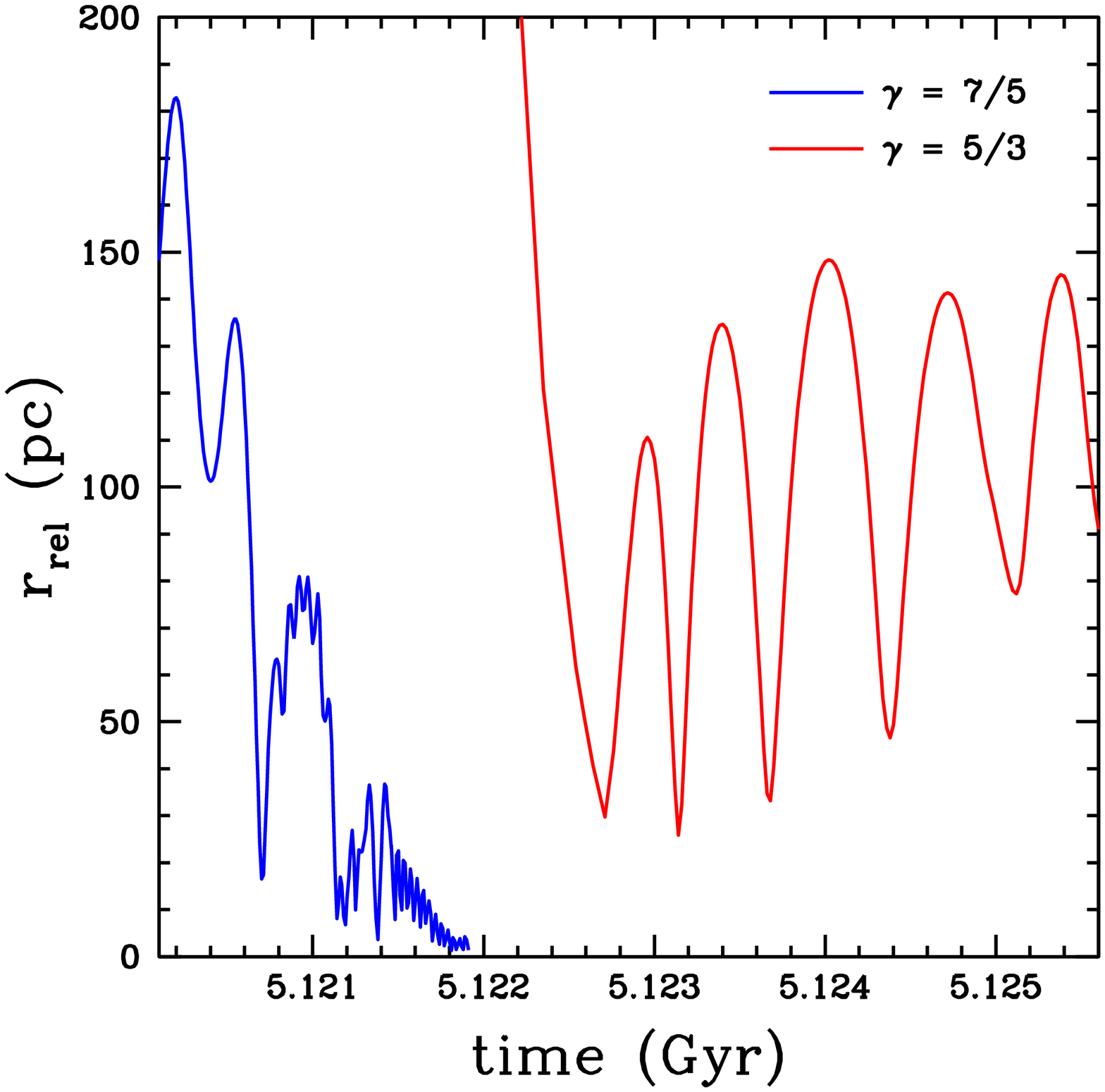}
\end{center}
\caption[]{Orbital separation of the binary BHs.  The blue line shows
the relative distance as a function of time for $\gamma=7/5$ as shown
in Figure 2, while the red line shows it for $\gamma=5/3$ }
\label{fig:stall}
\end{figure}

We tested how a smaller degree of dissipation in the gas affects the
structure and dynamics of the nuclear region by increasing $\gamma$ to
$5/3$. This would correspond to a purely adiabatic evolution. The radiative
injection of energy from an active nucleus  is a good
candidate for a strong heating source that our model does not take
into account (Spaans \& Silk 2000; Klessen, Spaans \& Jappsen 2007).
An AGN would not only act as a source of radiative heating but would
also increase the turbulence in the gas by injecting kinetic energy
(Springel, Di Matteo \& Hernquist 2005) in the surrounding medium,
possibly suppressing gas cooling.  Before the two galaxy cores merge,
double (or single) AGN activity can in principle alter the thermal
state of the gas.

We have run another refined simulation with $\gamma=5/3$ to explore
this extreme situation. In this case we find that a turbulent,
pressure supported cloud of a few hundred parsecs arises from the
merger rather than a disc. The mass of gas is lower within 100 pc
relative to the $\gamma=7/5$ case because of the adiabatic expansion
following the final shock at the merging of the cores. The nuclear
region is still gas dominated, but the stars/gas ratio is $> 0.5$ in
the inner $100$ pc.  
This suggests that the 
$\gamma=5/3$ simulation does not describe the typical nuclear
structure resulting from a dissipative merger.

The BH duet does not form a binary owing to inefficient orbital decay,
and maintains a separation of $\sim 100-150$ pc, as shown in Figure 4.  
The gas is hotter and more turbulent; the sound speed
$c_{\rm s}\sim 100$ km s$^{-1}$ and the turbulent velocity $v_{\rm turb}
\sim 300$ km s$^{-1}$ are of the same order of $v_{\rm BH}$, the
velocity of the BHs, and the density around them is $\sim 5$ times
lower than in the $\gamma=7/5$ case.  Stars and gas will drive the BHs
closer to form a binary, but on an estimated dynamical friction time
of several $10^7$ yr (Mayer \etal 2007, supporting online material).

\section{BH inspiral in equilibrium rotating nuclear discs}

\subsection{Initial conditions}

In this section, we present an independent series of simulations
(carried out with {\sl GADGET}; Springel, Yoshida \& White 2001) that
trace the dynamics of a BH pair (with $q_{\rm BH}=1, 1/4, 1/10$)
orbiting inside a self-gravitating, rotationally supported disc
composed either of gas, gas and stars, or just stars (Dotti, Colpi \&
Haardt 2006; Dotti \etal 2007).  The main parameters of the
simulations are summarized in Table 1.

\begin{table}\label{tab:run}
\begin{center}
\caption{Run parameters}
\vskip 0.5 cm
\begin{tabular}{l@{   }c@{   }c@{   }c@{   }c@{   }c@{   }c@{   }c@{       }c@{   }c@{   }}  
\hline
run & central BH & $f_*^a$ & $M_{\rm BH,1}^b$ & $M_{\rm BH, 2}^b$ & $M_{\rm Disc}^b$  & 
$M_{\rm Bulge}^b$ & $e$ &  Q & res$^{c}$ \\
\hline
\hline 
A1    &    &   &   &   & 100 &     & 0    &      &        \\
A2    & no   & 0 & 1 & 1 & 100 & 698 & 0.9  &  1.8 & 1 \\
A3    &    &   &   &   & 0    &     & 0.9 &      &        \\ 
\hline
B1    &    &   &   &   &     &     & 0    &      &        \\
B2    & no   & 0 & 5 & 1 & 100 & 698 & 0.9  & 1.8  & 1 \\
B3$^{d}$&  &   &   &   &     &     & 0.9  &      &        \\
\hline 
C1$^{e}$    &    &   &   & 4   &     &     &     &     &        \\
C2    &  yes  & 0 & 4 & 1   & 100 &  698 & 0.7& 3   & 1 \\
C3    &    &   &   & 0.4 &     &     &  &        &        \\ 
\hline
D1    &    &   &    & 4     &     &     &  &     &        \\ 
D2    &  yes  & 1/3& 4 & 1    & 100 & 698 & 0.7 & 3 & 1 \\
D3    &    &   &   & 0.4    &     &     &  &     &        \\
\hline
E1    & &   &    & 4   &     &     & &                &       \\
E2    & yes & 2/3& 4 & 1    & 100 & 698 & 0.7 & 3         & 1  \\
E3&   & &   & 0.4    &     &     & &              &        \\
\hline  
F1  & &   &   &   4  &     &     & &                &        \\
F2  & yes & 1 & 4 & 1    & 100 & 698 & 0.7 & 3           &  1  \\
F3  & &   &   & 0.4   &     &     & &               &        \\
\hline 
G1$^{f}$& yes & 0 & 4 & 4 &  100  & 698 & 0.4 &  3  & 0.1    \\
\hline
\end{tabular}\\
\end{center}
\noindent
\footnotesize{$^{a}$ $f_*$: disc mass fraction in stars.}\\
\footnotesize{$^{b}$ BH masses are in units of $10^6 \,\msun$.}\\
\footnotesize{$^{c}$ Force resolution in pc.}\\
\footnotesize{$^{d}$ The secondary lighter 
BH in run B3 has a retrograde orbit.}\\
\footnotesize{$^{e}$ simulation C1 was run two times, 
setting the secondary BH on a prograde and a
retrograde orbit.}\\
\footnotesize{~$^{f}$ Simulation C1 re--run using the  
particle splitting technique to improve force resolution.}\\
\end{table}

In all the simulations there are two BHs, a stellar bulge, and a
massive rotationally supported nuclear disc composed by stars, gas, or
both (except A3).  The disc has a mass $M_{\rm Disc}=10^8 \msun$, an
extension of 100 pc, a vertical thickness of 10 pc, and follows a
Mestel surface density profile: \be
\label{eq:mestel} \Sigma_{\rm Disc} (R)= \frac{K_{\rm Disc}}{R}, \ee
where $R$ is the radial distance projected into the disc plane, and
$K_{\rm Disc}$ is determined by the total mass of the disc.  The
vertical profile of the disc is initially set homogeneous.  The
rotational velocity of the gas $v_{\rm rot}$ is constant through the
disc and this implies that fluid elements are rotating differentially
with an angular velocity $\Omega_{\rm rot}=v_{\rm rot}/R.$ The
spheroidal stellar bulge is modeled with $10^5$ collisionless
particles, initially distributed as a Plummer sphere with mass density
profile: \be \rho_{\rm Bulge} (r)={3 \over 4 \pi}{\mbulge\over b^3}
\left(1+{r^2\over b^2}\right)^{-5/2}, \ee where $b$ $(=50$ pc $)$ is
the core radius, $r$ the radial coordinate, and $\mbulge(=6.98
M_{\rm{Disc}})$ the total mass of the spheroid.  With such choice, the
mass of the bulge within $100$ pc is 5 times the mass of the disc, as
suggested by Downes \& Solomon (1998).  We relax our initial composite
model (bulge, disc and, if present, the central BH) for $\approx 3$
Myrs, until the bulge and the disc reach equilibrium. Given the
initial homogeneous vertical structure of the disc, the gas initially
collapses on the disc plane exciting small waves that propagate
through the system.  The parameters varying in the simulations are:

\noindent
$\bullet$ The disc mass fraction in stars.
We run four different sets of simulations assuming
a purely gaseous disc (runs A, B, C, and G),
and a disc in which 1/3 (runs D), 2/3 (runs E), and,
finally, all gas particles (runs F) are turned into collisionless
particles, respectively. For each disc model with fixed star
fraction, we evolved the initial condition in isolation until
equilibrium is reached.
We do not convert any gaseous
particle in stars when we follow the dynamics of the BHs, so
that the disc stellar fraction remains constant;

\noindent
$\bullet$ The disc mass. In all the simulations the disc mass is
M$_{\rm Disc} = 10^8 \msun$, apart from run A3, in which we study the
BH pair evolution in a pure stellar bulge.  

\noindent
$\bullet$ 
The Toomre parameter $Q$ of the disc. In our simulations with
a pure gaseous disc (runs A, B, C, and G), we set a initial internal energy
profile $u(R) = K_{\rm Th}~ R^{-2/3}$, where $K_{\rm Th}$ is a constant
defined so that the Toomre parameter of the disc,
\be \label{toomre}
Q=\frac{k c_{\rm s}}{\pi G \Sigma}
\ee
is $\gsim 1.8$ for runs A and B (cold disc runs), and $\gsim 3$ for runs C
and G (hot disc runs) everywhere.
In equation~(3.3), $k$ is the local epicyclic frequency and
$c_{\rm s}$ is the local sound speed of the gas. The internal energy of
the gas is evolved adiabatically neglecting radiative cooling/heating
processes. Our choice of $Q>$ 1.8 everywhere prevents fragmentation
of the disc, and, when $Q>$ 3 formation of large--scale
over-densities, such as spiral arms or bars. We do not model any turbulent
motion in the disc, but we consider the internal energy as a form of unresolved
turbulence, and, as a consequence, $c_{\rm s}$  as local turbulent velocity. 
In the simulations with a non--null stellar
fraction of the disc mass (runs D, E, and F) we set the local velocity
dispersion of the disc stars equal to the local sound speed.
With this procedure $Q$ is the same than in a pure
gaseous disc.   

\noindent
$\bullet$ 
The masses of the two BHs. As shown in Table 1, we explore
various mass ratios $q_{\rm BH}$ from 1 to 1/10.

\noindent
$\bullet$ The BH orbits. The BHs are placed on coplanar orbits that
can either be prograde or retrograde, circular or eccentric (with
$e$=0.4,0.7,0.9).  In run A and B the two BHs are placed at 55 pc from
the dynamical center of the disc. In run C,D,E,F the primary BH is at
the center of the disc.

\noindent
$\bullet$
The initial eccentricity of the orbiting BH. In all the runs
of A and B classes a BH is initially moving on a circular orbit, while the
second BH can have an orbital eccentricity ($e$) of 0 (circular motion,
runs A1 and B1) or 0.9. In runs C, D, E, F, and G a BH is initially
placed at rest in the center of the structure (see point 1) while the
orbiting BH is initially moving on eccentric orbits with $e=$ 0.7
in runs C, D, E, and F, $e=$ 0.4 in run G1;

\noindent
$\bullet$ The spatial resolution of the simulations. In our low
resolution simulations (runs A, B, C, D, E, and F), we model our
relaxed Mestel disc with 235331 particles, and use a number of
neighbors of 50.  This number defines a subsample of particles used by
the code to evaluate the local hydrodynamical parameters. This number
define, in the center of the disc, a hydrodynamical force resolution
(usually defined as smoothing length) of $\approx 1$ pc. We set also
equal to 1 pc our gravitational softening, the parameter setting the
spatial resolution of the gravitational acceleration evaluation.
Evaluating the two forces, hydrodynamical and gravitational, with the
same resolution prevents spurious local condensation of gas ( occurring
when the smoothing length exceeds the gravitational softening) or
outflows (in the opposite case). The stellar gravitational softening
is set equal to the gaseous one.  In our high resolution simulation
(run G1) we re-sample the output of simulation C1 thanks to the
technique of particle splitting (Kitsionas \& Whitworth 2002). With
this technique we locally increase the number of gaseous particles of
one order of magnitude reducing accordingly the smoothing length and
the gravitational softening.  The new spatial resolution in the
central regions is $0.1$ pc.

\subsection{Orbital decay on circular orbits}

\begin{figure}
\includegraphics[width=0.495\textwidth]{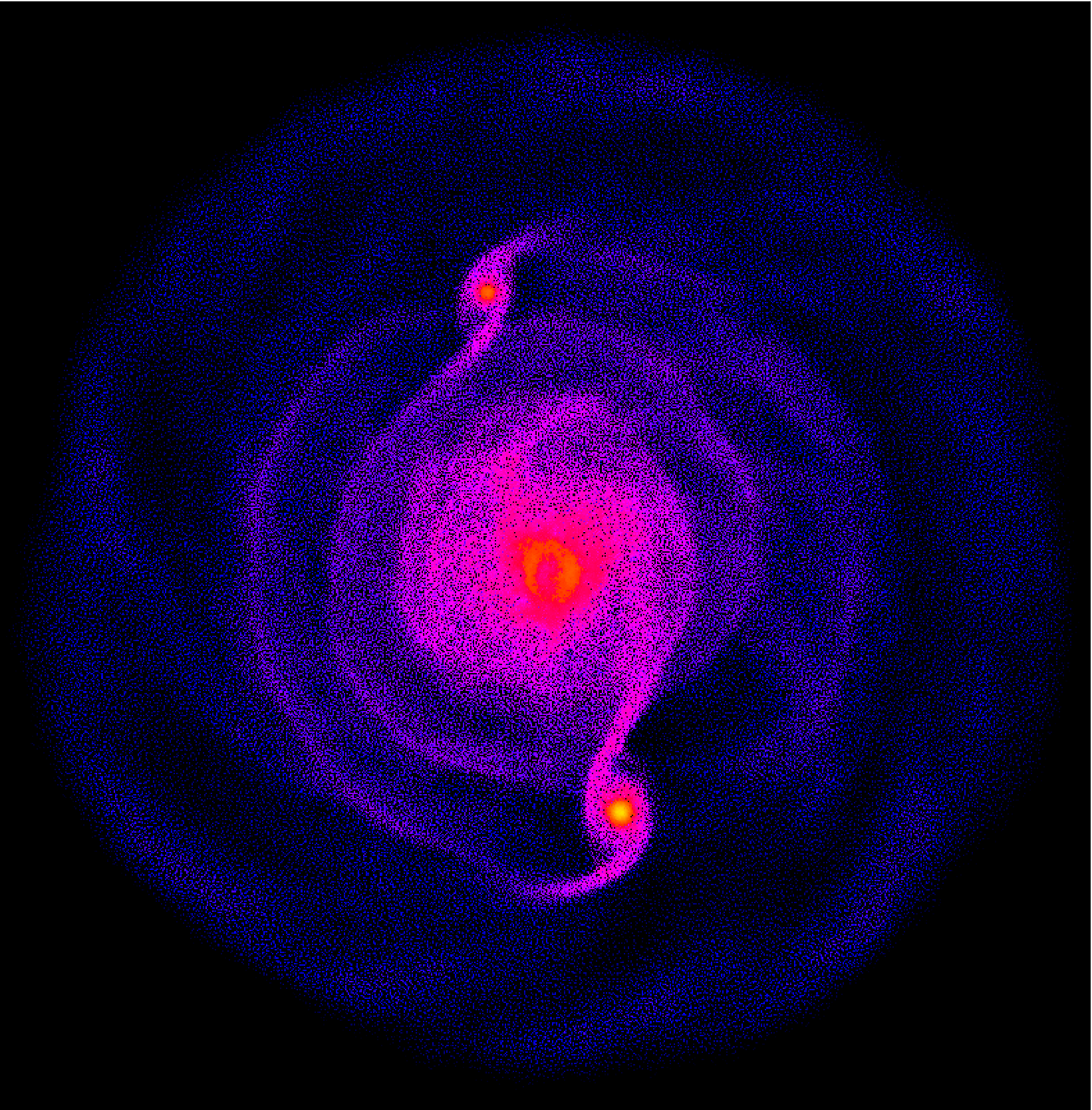}
\includegraphics[width=0.505\textwidth]{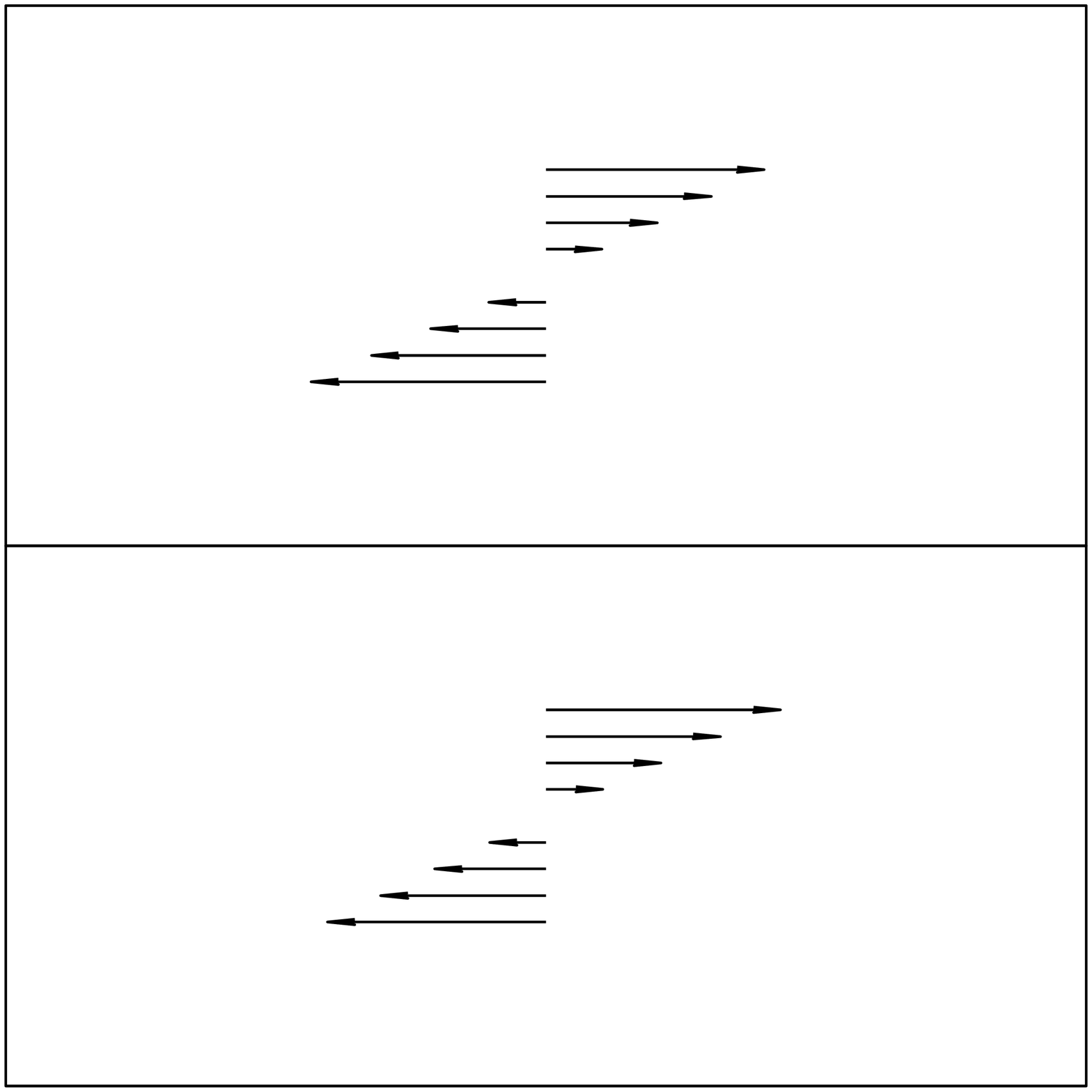}

\caption{Left panel:  face--on projection  of the disc  for run  B1 at
time 1 Myr. The color coding  shows the z--averaged gas density on a
logarithmic scale  between 100 (dark blue) and  3$\times\, 10^5\, {\rm
M}_{\odot}\,  {\rm pc}^{-3}$  (yellow).  The two  orbiting bright  dots
highlight the position of the two BHs (the top BH is the lighter one)
theta are moving counterclockwise. They  excite prominent wakes along
their trails.
Right panels:  relative angular velocity pattern between  the BHs and
the gaseous disc for different radii. The upper(lower)--right panel is
centered in the  upper (lower) BH radius. The  separations between two
arrows are 1 pc.
}\label{fig:-001}
\end{figure}

In simulations A1 and B1, the BHs
move initially on circular
prograde orbits inside the disc. 
The initial separation, relative to the center of mass, is 
$\approx 55$ pc.

We plot in Figure 5 the density map of the gas surrounding the unequal
mass BHs in run B1 at a selected time. Both BHs are exciting prominent
density wakes whose extent depends on the amount of disc mass
perturbed by the orbiting BHs, which is a function of the BH masses,
as can be noted in the Figure.  The presence of two wakes near each BH
is due to the different angular velocity pattern of the gas in the
disc.  Given our choice for the bulge and disc profile, in the BH
comoving frame, the disc inside the BH orbit is moving
counterclockwise while the outer regions are moving clockwise, as
shown in the right panel of Figure~5. In a non rotating frame, two
wakes develop around each BH, one outside the orbit and one inside.
The motion of the BH is highly supersonic, and this explains the
coherent structure and shape of their wakes (Ostriker 1999).  The
interaction between the BH and it inner wake increases the BH angular
momentum, while the outer brakes the motion. Given our choice of the
density/internal energy profiles, and the velocity field in the disc,
the outer wake is more effective in removing orbital energy and
angular momentum from each BH compared to the positive (accelerating)
torque exerted by the wake excited in the inner part of the disc. The
net effect is the braking of the BH pair, so that a binary can form on
a timescale of $\sim 10^7$ yr.

\subsection{Orbital decay along eccentric orbits}

\begin{figure}
\begin{center}
\centerline{\psfig{figure=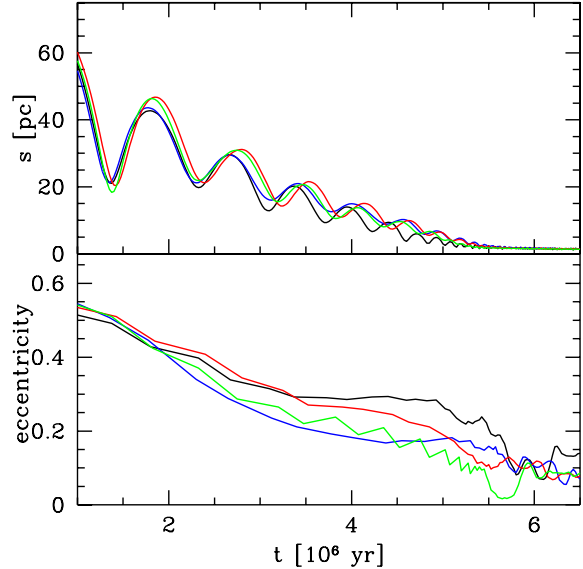,height=8cm}}
\caption{Equal mass BHs. Upper panel: separations $s$ (pc) between the
BHs as a function of time. Lower panel: eccentricity of the BH binary as
a function of time. Black, blue, red and green lines refer to stellar
to total disc mass ratio of 0, 1/3, 2/3 and 1 (run C1, D1, E1, and F1) 
respectively.
}
\label{fig:000} 
\end{center}
\end{figure}

Figure~6 shows the BH relative separation $s$ and eccentricity as a
function of time (upper and lower panel respectively) for equal mass,
initially eccentric binaries of runs C1, D1, E1, and F1, where a
primary BH is initially at rest in the dynamical center of the disc.
Regardless the fraction of star--to--gas disc particles, the secondary
BH spirals in, at the same pace, reaching $\sim 1$ pc after $\sim 5$
Myr.  The velocity dispersion of the disc--stars is similar to the gas
sound speed, and the two components share the same differential
rotation. This is why dynamical friction on the secondary BH, caused
by stars and gas, is similar.  As the orbit decays, the eccentricity
decreases to $e\lsim 0.2$.  This value is not a physical lower limit,
but rather a numerical artifact due to the finite resolution, as will
be shown in run G1.

Here, we show that circularization occurs regardless the nature of 
the disc particles (gas and/or stars). Note that circularization
takes place well before the secondary feels the gravitational
potential of the primary, so the BH mass ratio does not play any
role in the process.

\begin{figure}
\begin{center}
\centerline{\psfig{figure=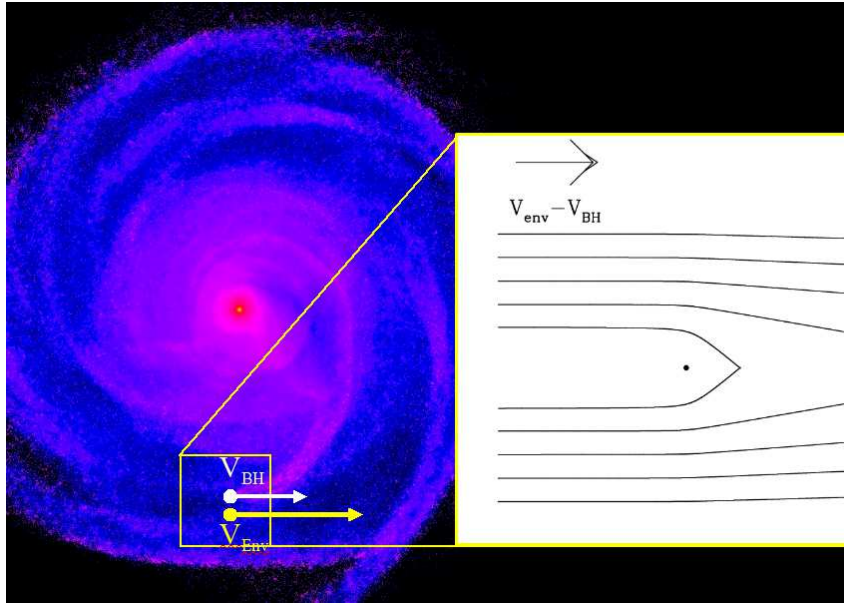,height=8cm}}
\caption{Face--on projection of the disc for run C1 at the first
(co-rotating) BH apocentre. The color coding shows the z--averaged
gas density, the white and yellow arrows refers to the BH and
disc velocities, respectively. In the insert panel 
the trajectories of the gas particles
are drawn, as observed in a frame comoving with the orbiting
BH. The density wake is in front of the BH trail.}
\label{fig:001} 
\end{center}
\end{figure}

To show how the circularization process works, let us consider two
different snapshot of run C1. In Figure~7 we plot the gas
densities in the disc at the time corresponding to the first passage
at the apocentre. In this simulation, the initially orbiting BH
is corotating with the disc, with a speed equal, in modulus, to the
velocity corresponding to a circular orbit at that initial position, 
but with an orbital eccentricity of 0.7. Near the
apocenter the BH is moving slowlier than the gas in the disc, as
highlighted by the two vectors in Figure 7, and in a frame comoving
with the BH, gas blows ahead of the BH.
The gas average rotational velocity decreases due to the 
gravitational interaction with the BH, and 
the back--reaction on the BH is a temporary 
 ``increase'' of its orbital energy
and angular momentum. This excites a wake in the forward direction,
i.e., at the apocenter the density wake is in front of
the BH, as shown in Figure~\ref{fig:001}.

In Figure~8 we plot the gas densities in the disc at the time
corresponding to the first passage at the pericenter.  The BH has
there a velocity higher than the local rotational velocity, so that
dynamical friction causes a drag, i.e. a reduction of the BH velocity.
Now, a wake of particles lags behind the BH trail.  Thus, given this
velocity difference, the wake reverses its direction decelerating
tangentially the BH.  Now the wake is behind the BH direction of
motion.  The orbital energy decreases more effectively than angular
momentum, since at pericenter the gas density is also higher than at
apocenter.  The overall effect is a net circularization of the orbit.
This seems a generic feature of dynamical friction, regardless the
disc composition, and holds as long as the rotational velocity of the
gas or/and star particles exceeds the gas sound speed and the stellar
velocity dispersion, as in all cases explored.  We remark that in
spherical backgrounds, dynamical friction tends to increase the
eccentricity, both in collisionless (Colpi, Mayer \& Governato 1999;
van den Bosch \etal 1999; Arena \& Bertin 2007) as well as in gaseous
(Sanchez--Salcedo \& Brandenburg 2001) environments.

\begin{figure}
\begin{center}
\centerline{\psfig{figure=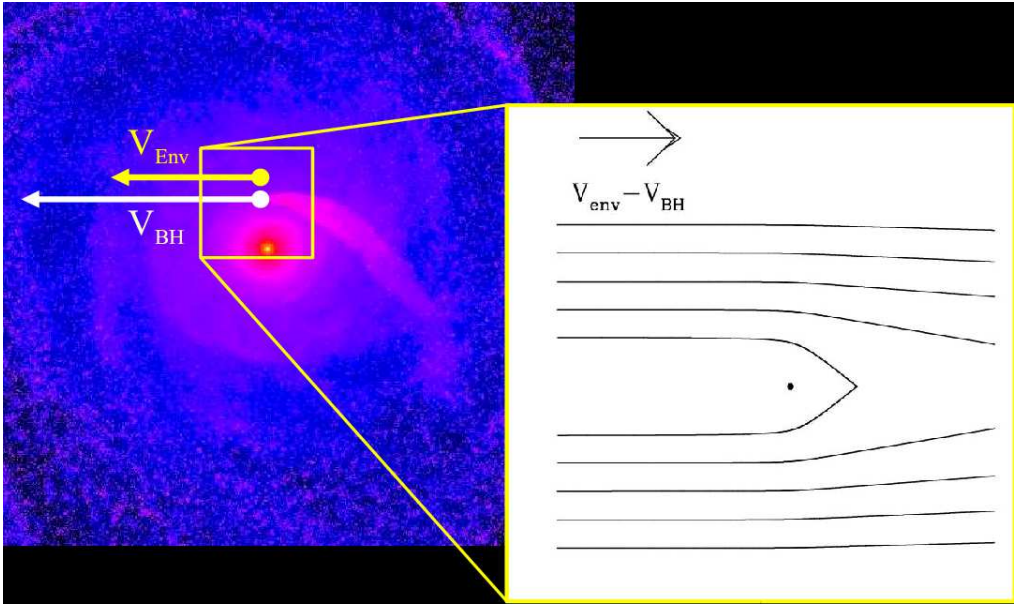,height=8cm}}
\caption{Same as Figure~7, but when the BH is 
at its first pericenter. The density wake is behind the BH trail.
}
\label{fig:002} 
\end{center}
\end{figure} 

The interaction between the BH and the disc for
counter--rotating orbits (run B3) keeps the BH orbit eccentric.
When a counter--rotating BH is near apocenter, the gas
flows against the BH motion.  The 
over-density that forms is behind the BH trail, and this occurs also near
pericenter. Dynamical friction
is weaker than for
co-rotating orbits
given the larger relative velocity between the BH and the gas; thus
the corresponding decay timescale is longer (by a factor $\sim 2$).


\subsection{High resolution run: dynamics}

We run a higher resolution simulation to study the eccentricity and
orbital evolution on scales smaller than 1 pc. The new
initial condition is obtained re-sampling the output of run C1 (for equal mass
BHs) with the technique of particle splitting. Re-sampling is performed when the
BH separation is $\simeq 14$ pc (corresponding to $\simeq 4$ Myr after
the start of the simulation). Splitting is applied to all
particles whose distance from the binary center of mass is $\leq 42$
pc, so that the total number of particles increases only by a
factor $\simeq 4$, while the local mass resolution in the split region
is comparable to that of a standard $\simeq 2 \times 10^6$ particle
simulation with uniform resolution. Our choice of the maximum 
distance for splitting is conservative, since it is aimed at preventing that more massive,
unsplit gas particles reach the binary on a timescale shorter than
the entire simulation time. 
In the central split region, the high mass resolution achieved fulfills
the Bate \& Burkert (1997) criterion for gravitational softening values
down to $0.1$ pc.

In Figure~9 we compare the surface density profile of the
circum--nuclear gaseous disc in run C1 at $t=4$ Myr, for the low and 
high resolution cases. The two profiles differ only below  
the scale of the low resolution limit $R\lsim 3$ pc. 
The decrease of the gravitational softening corresponds
to introducing a deeper potential well of the BH
within a sphere defined by the former softening radius. 
Therefore, with the improved resolution, the central
surface density increases as the gas reaches a new
hydrostatic equilibrium closer to the BH, as shown
in Figure~9 (red line).
The lack of noticeable differences in the surface profile at
separations $R>3$ pc confirms the accuracy of the particle splitting
technique.

\begin{figure}
\begin{center}
\centerline{\psfig{figure=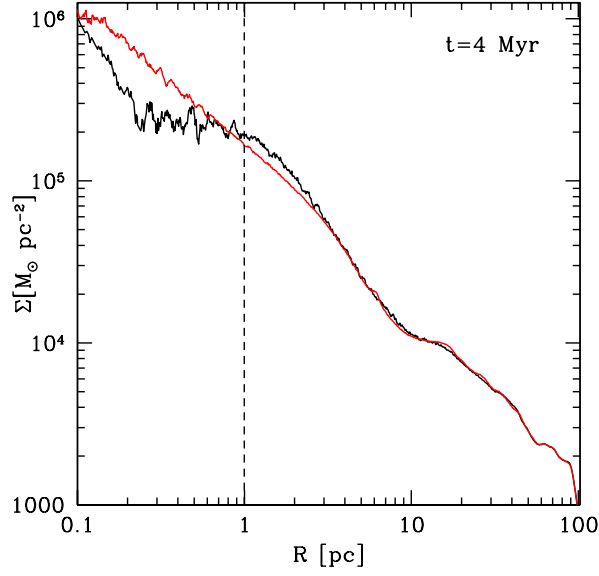,height=8cm}}
\caption{Surface density profile of the circum--nuclear gaseous disc in run
C1 at $t=4$ Myr. Black line refers to the surface density in the low
resolution simulation, red line refers to the high resolution (split)
simulation. The dashed vertical line marks the resolution limit in the
un--split simulation.
}
\label{fig:004} 
\end{center}
\end{figure} 

Results of the high resolution run are shown in Figure~10.
The separation decays down to $0.1$ pc in $\sim 10$ Myr.
In the high resolution run, the dynamical
evolution of the BHs is initially identical to the low resolution case. 
Because of particle
splitting, the system granularity is reduced, 
and therefore the force resolution increases. 
In the high resolution run, the binary
decreases its eccentricity to $\approx 0$ (before
the new spatial resolution limit is reached).

\begin{figure}
\begin{center}
\centerline{\psfig{figure=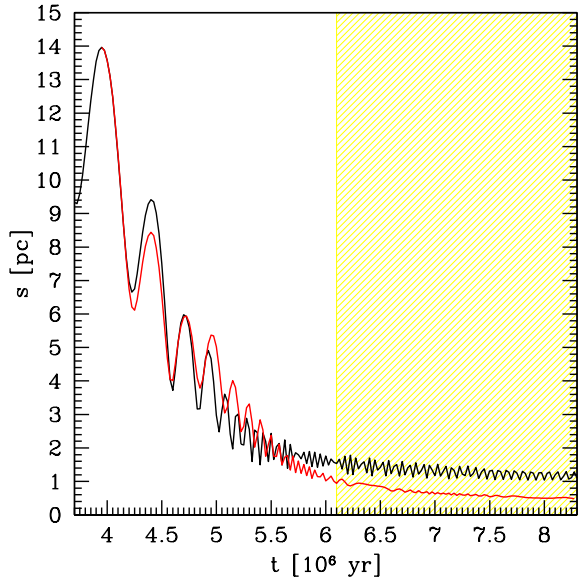,height=8cm}}
\caption{
  Separations $s$ (pc) between the BHs as a function of time.  Red
  (black) line refers to the (un--)split run C1. The dashed area
  corresponds to the region where the BH separation is $<1$ pc in the
  split higher resolution.  }
\label{fig:005} 
\end{center}
\end{figure} 

\subsection{High resolution runs: constrains on accretion processes}

In run G1, a resolution of 0.1 pc allows us to study the properties of
the gas bound to each BH.  To this purpose, it is useful to divide
gaseous particles, bound to each BH, into three subsets, according
to their total energy relative to the BH. We then define weakly
bound (WB), bound (B), and strongly bound (SB) particles according to
the following rule: \be \label{bound} E < \left\{
\begin{array}{ll}
0  & (\mbox{WB}) \\
0.25\, W & (\mbox{B}) \\
0.5 \,W  & (\mbox{SB}),  
\end{array}
\right.  \ee where $E$ is the sum of the kinetic, internal and
gravitational energy (per unit mass), the latter referred to the
gravitational potential $W$ of each individual BH.  Hereafter WBPs,
BPs, and SBPs will denote particles satisfying the WB, B, or SB
condition, respectively.  Note that, with the above definition, SBPs
are a subset of BPs, which in turn are a subset of WBPs.

We find that the mass collected by each 
BH, relative to WB, B, and SB particles is
$M_{\rm WBP}\approx 0.85 M_{\rm BH}\approx 3.4 \times 10^6 \msun$,
$M_{\rm BP}\approx 0.41 M_{\rm BH}\approx 1.6 \times 10^6 \msun$, and 
$M_{\rm SBP}
\approx 0.02 M_{\rm BH}\approx 8 \times 10^4 \msun$, respectively 
(here $M_{\rm BH}=M_1=M_2$).
These masses are remain constant 
with time as long as the BH separation is $s \gsim 1$ pc. 
At shorter separations WBPs and BPs are perturbed by the tidal field 
of the BH companion, and at the end of the simulation $M_{\rm WBP}$ and 
$M_{\rm BP}$ are reduced of a factor $\approx 0.1$.
During the same period of time, $M_{\rm SBP}$ associated to the 
primary (secondary) BH increases
by a factor $\approx 4$ ($\approx 2.5$). 
This result is unaffected by numerical noise 
since the number of bound particles (associated to each class) 
is $\gsim 1$ SPH kernel ($N_{\rm neigh} = 50$). 

The radial density profiles of WBPs, BPs, and SBPs are well resolved
during the simulation. Bound particles 
have a net angular momentum with respect to each BH, and form a
pressure supported spheroid. 
The half--mass radius is similar for the two BHs: 
$\simeq 3$ pc, $\simeq 1$ pc, and $\simeq 0.2$ pc for WBPs, BPs,
and SBPs, respectively.
The disc gas density can be as high as $10^7$ cm$^{-3}$. It is
then conceivable that, at these high densities, dissipative processes could
be important, possibly reducing the gas internal (turbulent and
thermal) energy well below the values adopted in our simulations.
If cooling becomes effective, we expect that the bound gas will form a 
geometrically thin disc with Keplerian angular momentum
comparable to what we found in our split simulation.
Since $L_{\rm z}=\sqrt{G\,M_{\rm BH}\, R_{\rm BH,disc}}$, we obtain,
for the primary BH, an
effective radius $R_{\rm BH,disc} \approx 0.1$ pc ($0.03$ pc) for WBPs (BPs).
The secondary BH is surrounded by particles with a
comparatively higher angular momentum
with  a corresponding effective radius $R_{\rm BH,disc}\approx 1$ 
(0.13) pc for WBPs (BPs). Finally, for SBPs, both BHs have 
$R_{\rm BH,disc} \ll 0.01$ pc, which is more than
an order of magnitude below our best resolution limit. These simple
considerations indicate that a more realistic treatment of gas
thermodynamics is necessary to study the details of gas accretion onto
the two BHs during the formation of the binary, and the
subsequent orbital decay. Nonetheless, our simplified treatment
allows us to estimate a lower limit to the accretion timescale,
assuming Eddington limited accretion: 
\be t_{\rm
  acc}=\frac{\epsilon}{1-\epsilon}\,t_{\rm Edd} \,
\ln{\left(1+\frac{M_{\rm acc}}{M_{{\rm BH,}0}}\right)}, 
\ee 
where $\epsilon$ is the radiative efficiency, 
$t_{\rm Edd}=c \sigma_T/(4\pi Gm_{\rm p})$ the
Salpeter time, $\sigma_T$ the Thompson 
cross section, $M_{{\rm BH,}0}$  the initial BH mass, and
$M_{\rm acc}$  the accreted mass. 
Assuming $\epsilon=0.1$, Eddington limited accretion can last
for 
$\sim 15$ Myr, and only $\lsim1$ Myr, if the BHs accrete
all the BPs, and SBPs, respectively.

Figure 11 depicts, in a cartoon, the configuration of the two
accretion discs surrounding the BHs inside the grand disc.  It is
expected that the two discs will eventually touch and disrupt tidally,
and re-organize to form a circum--binary geometrically thin
disc surrounding both BHs; exploring this configuration is the goal of
our next series of simulations.  Further braking of the BH motion and binary
hardening requires energy loss and angular momentum transport through
a mechanism that may be reminiscent of planet migration in
proto-stellar discs (Gould \& Rix 2000): while gravitational torques
carry away angular momentum, viscous torques inside the cool
Keplerian circum--binary disc sustain the radial motion of the gas 
toward the BHs
maintaining the binary in near contact  with the disc. 
Equilibrium between the gravitational 
and viscous torques cause the slow drift of the
BHs toward smaller and smaller separations until the gravitational
waves guide final inspiral. 

\begin{figure}
\begin{center}
\centerline{\psfig{figure=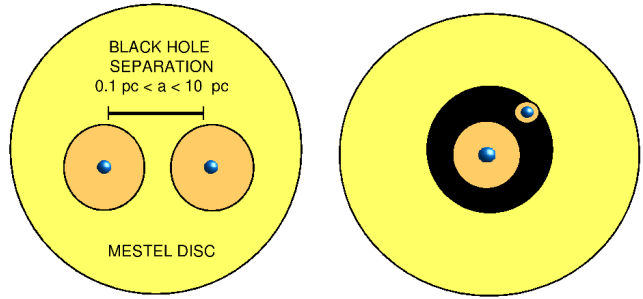,height=10cm,angle=270}}
\caption{
Left: a cartoon depicting the BH binary when the BH separation is
$\gsim 0.1$ pc, as suggested by our high-resolution simulation. The
BHs are surrounded by their own accretion discs. Further orbital
decay by gravitational and viscous torques are expected to 
open a gap leading to the configuration depicted in the right panel.}
\label{fig:005} 
\end{center}
\end{figure}

\section{BH binaries, the $M_{\rm BH}$ versus $\sigma$ relation and the stalling problem}

In the previous section, we followed the orbital decay of a BH binary
in a rotationally supported gaseous background, down to a scale of
$\sim 0.1$ pc, and found that the hardening of the BH binary, and thus
the corresponding energy and angular momentum loss, results from 
large--scale density perturbations excited by the BH gravitational.
These perturbations have still negligible impact on the overall
structure of the Mestel disc: compared with the binding energy of the
disc, the energy deposited by the BH binary is still small ($\approx$
an order of magnitude lower).  But this may not hold true if the
binary continues to harden down to the scale where gravitational wave
emission becomes important.  Unless gravitational torques are capable
of extracting angular momentum with no energy dissipation (making the
binary more and more eccentric until $e\to 1$), the energy input from
the BH binary may have two main effects: (i) to modify the thermal and
density structure of the grand disc; (ii) to halt the BH binary
hardening (by ``evaporating'' the environment) before gravitational
wave emission intervenes to guide the BHs toward coalescence.  This
``negative'' feedback on the BH binary fate may indeed cause the
``stalling'' of the binary, a problem that has been discussed mainly
in the context of pure stellar backgrounds (e.g. Merritt 2006a).

In this section we would like to introduce a simple argument based on
energy conservation (i.e., assuming that angular momentum transport is
accompanied by some energy dissipation) in the attempt to investigate whether
the BH binary deposits enough energy to modify its surroundings and
influence its fate.

\begin{figure}
\begin{center}
\includegraphics[height=0.50\textheight]{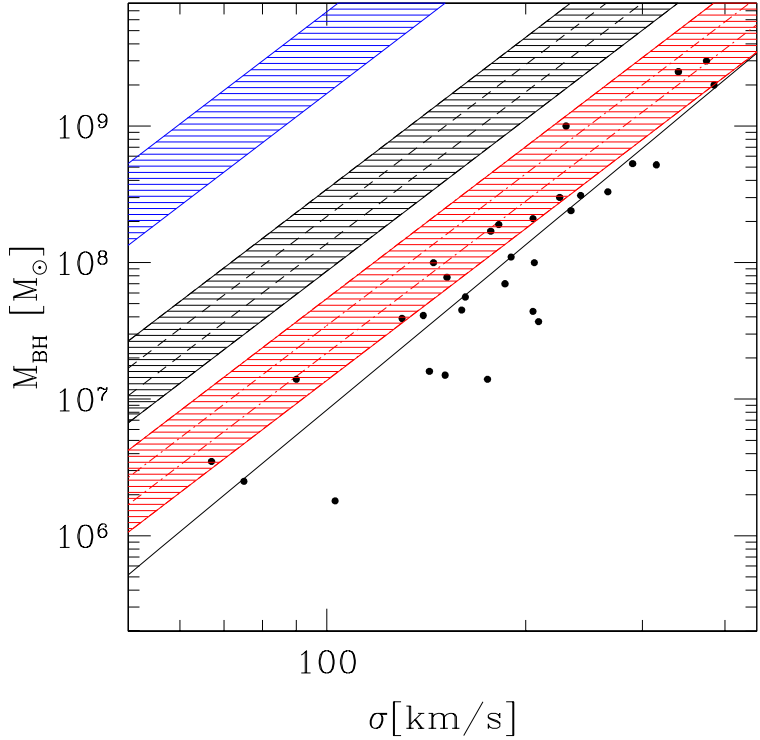}
\end{center}
\caption[]{
  $M_{\rm BH}$--$\sigma$ plane: thin solid line is the best fit to the
  data (Tremaine et al. 2002). Black (blue) lines refer $M^{\max}_{\rm
    BH}\equiv {\cal {M}}(t_{\rm GW},e,q_{\rm BH})\,\sigma^{3.7}$ (see eq.
  [4.8]) computed for $\alpha=2.63$, 
$q_{\rm BH}=1$, $e=0$ ($e=0.99$) and $t_{\rm
    GW}=10^9-10^8-10^7-10^6$ yr (from top to bottom).  The red strip
  is computed using equation (4.8) for only the gaseous
  component, assuming $E_{\rm iso,gas}=0.1\,E_{\rm iso,bulge}$.}
\label{fig:stall}
\end{figure}

Thus, consider a BH binary of total mass $M_{\rm BH},$ reduced mass
$\mu_{\rm BH},$ and mass ratio $q_{\rm BH}$.  The binary will
coalesce in a time $t_{\rm GW}$ under the action of gravitational wave
emission if its semi-major axis $a$ is smaller than 
\begin{equation}
a_{\rm GW}=\left ( {256\over 5}{G^3\over c^5} f(e)  M_{\rm BH}^2\mu_{\rm BH} t_{\rm GW}\right )^{1/4}, 
\end{equation}
where $f(e)=[1+(73/24)e^2+(37/96)e^4](1-e^2)^{-7/2}$ 
(note that as $e\to 1,$ $a_{\rm GW}\to \infty$). 
This separation is of only  $8\times 10^{-4}$ pc if we consider 
a BH binary of total mass $10^6\msun,$ with eccentricity $e=0$,
mass ratio $q_{\rm BH}=1$, and merging time of $t_{\rm GW}=10^9$ yr.

If the BH are initially unbound (as it is the case of a galaxy merger) 
the total energy $\Delta E_{\rm BBH}$  that the BH binary
needs to deposit 
in its environment to reach $a_{\rm GW}$ is 
\begin{equation}
\Delta E_{\rm BBH}={1\over 2}
\left ({256\over 5} {G^3\over c^5}f(e)\right)^{-1/4}G M_{\rm BH}^{5/4}
{q_{\rm BH}^{3/4}\over (1+q_{\rm BH})^{3/2}} t_{\rm GW}^{-1/4}.
\end{equation}
For the binary considered

\begin{equation}
\Delta E_{\rm BBH}\approx 1.3\times 10^{55}\left (
{M_{\rm BH}\over 10^6\msun}\right )^{5/4} \left ({10^9 \rm {yr}\over t_{\rm GW}
}\right )^{1/4}\,\,\,\ \rm {erg.}
\end{equation}

If we simply model the binary environment with an isothermal sphere of
effective radius $R_{\rm e}$, mass $M_{\rm iso}$ and 1-D velocity
dispersion $\sigma$, BH coalescence imposes, as necessary condition
\begin{equation}\Delta E_{\rm BBH}<E_{\rm iso}, 
\end{equation}
where $E_{\rm iso}=M_{\rm iso}\sigma^2$ is the binding energy of the
isothermal sphere.  An energy $\Delta E_{\rm BBH}$ of the order of
$E_{\rm iso}$ would modify the equilibrium structure of the sphere and
halt binary decay. 
If we identify the sphere with the stellar ``bulge'' hosting the BH
binary, and write 
$M_{\rm iso,
  bulge}={\cal{K}}\sigma^{\alpha},$ 
binary coalescence is  possible
if
\begin{equation}
{1\over 2}{ G\mu_{\rm BH} M_{\rm BH}\over a_{\rm GW}} < {\cal {K}} \sigma^{\alpha+2}.
\end{equation}
Then, using the Fundamental Plane relation (Cappellari et al. 2006) and
the mass--to--light ratio as in Zibetti et al. (2002), the bulge mass can be expressed as 
\begin{equation}
M_{\rm bulge}\approx  1.54\times 10^{11}\left ({\sigma\over 200 \kms}\right )^{2.63}\msun,
\end{equation}
and an approximate estimate to the binding energy is 
\begin{equation}
E_{\rm iso,bulge}\approx  \times 10^{59}
\left ({\sigma\over 200 \kms}\right )^{4.63}\rm {erg}.
\end{equation}
Equation (4.5) combined with (4.2) then results in an upper
limit for the binary BH mass
\begin{equation}
M_{\rm BH}< M^{\rm max}_{\rm BH}\equiv 
{\cal {M}}(t_{\rm GW},e,q_{\rm BH} ) \sigma^{4(\alpha+2)/5};
\end{equation}
\noindent
above  $M^{\rm max }_{\rm BH}$ the energy deposited by the
binary in its hardening would become comparable to the binding
energy of the surrounding and coalescence would be halted.

$M^{\rm max }_{\rm BH}$ depends on $\sigma$ and, for our choice of
$\alpha$, $M^{\rm max }_{\rm BH}\propto \sigma^{3.7}$ while its
normalization $\cal M$ is fixed by the coefficient $\cal K$
and by the parameters intrinsic to the binary, i.e.,
$t_{\rm GW}$, $e$, and mass ratio $q_{\rm BH}$.  For $t_{\rm GW}$
between $10^6$ yr and $10^9$ yr, $e=0,$ and $q_{\rm BH}=1$, the values
of $M^{\rm max}_{\rm BH}$ are inferred from equations (4.6) and (4.8), 
and plotted in
Figure 12 in black colors.  In the plane $M_{\rm BH}$--$\sigma$  we
overlaid the BH masses from Tremaine et al. (2002) for a comparison
with the observations.  If the observed BH masses in the sample of
Tremaine are the result of a BH binary merger that led also to the
formation of the host elliptical, we then conclude that BH binary
coalescences did not affect the equilibrium structure of the bulges.  If
we consider very eccentric ($e=0.99$) equal mass binaries, equations
(4.6) and 
(4.8) leads to upper mass limits (depicted with the blue strip)
again well above the observed points, thus
providing less stringent constraints on the BH binary fate.
Unequal mass binaries move further upwards, on the left side of
the diagram.
  
The decaying BH binary have thus no effect on the
overall stellar
bulge. However, they can have some influence inside the bulge, 
in a region much larger than the gravitational
influence radius of a ``single'' BH, defined as   
\begin{equation}
r_{\rm BH}=GM_{\rm BH}/\sigma^2.
\end{equation} 
Since the mass within radius $r$ 
of an isothermal sphere scales as $M_{\rm iso}(r)=2\, r\,\sigma^2/G$
we can determine the ``radius of BH binary
gravitational influence'', $r_{\rm BBH},$ 
obtained by equating $\Delta E_{\rm BBH}$ to the energy
inside $r_{\rm BBH}$, i.e.,
$ 
\Delta E_{\rm BBH}=M_{\rm iso}(r_{\rm BBH})\sigma^2:
$
\begin{equation}
r_{\rm BBH}={1\over 4}
\left ({256\over 5} {G^3\over c^5}f(e)\right)^{-1/4}
{q_{\rm BH}^{3/4}\over (1+q_{\rm BH})^{3/2}} t_{\rm GW}^{-1/4}
\left ({G^2M_{\rm BH}^{5/4}\over \sigma^4}\right ). 
\end{equation}
This radius is considerably larger than $r_{\rm BH}$.  For a
$10^8\msun$ equal mass--circular BH binary
$r_{\rm BBH}/r_{\rm BH}\approx 25,$ 
assuming $\sigma\sim
200\kms.$
The above relation may
thus account for the size of the stellar core seen in the bright
elliptical galaxies as extensively discussed by Merritt (2006b).

If the bulge hosts a gaseous nuclear component, the energy released by
the binary during its path to coalescence may alter the equilibrium of
the surrounding gas and a delicate interplay between heating/cooling
of the gas and hardening of the binary may lead to a
``self-regulated'' evolution of the BH mass and the circum--binary gas.
Given the
expected widespread values of $E_{\rm iso,gas}$ for a nuclear gaseous
component in merging galaxies, due to diversities in the mass gas
content, thermodynamics and equilibrium end--states, we simply rescale
$E_{\rm iso, gas}=0.1E_{\rm iso,bulge}$ and plot the corresponding red
strip in Figure 12, for $e=0,$ $q_{\rm BH}=1$ and same interval of
merging times ($t_{\rm GW}=10^6-10^9$ yr).  
The strip now shifts to the right and
gets closer to the BHs observed along the 
$M_{\rm BH}$--$\sigma$ relation. For $t_{\rm GW}$
less than $10^{7}$ yr, there are uncloalesced BHs and BHs that may
have deposited enough energy to affect the surrounding gas.
This point will be addressed in future investigations. 

A further energy constraint may come from accretion. First notice that 
the gravitational wave time scale $t_{\rm GW}$ is here treated as a parameter.
However in the nuclear region of the galaxy its value is determined by the
mechanisms guiding the inspiral, i.e., material (viscous) and  
gravitational torques and energy dissipation via shocks 
and radiation.
We know that 
viscosity is the critical parameter at the heart of
BH binary hardening on sub--parsec scales 
(Armitage \& Natarajan, 2002), and at the heart
of accretion (fixing the magnitude of the mass transfer rate toward
the BHs).
As shown in Section 3.4, gas inflows, after orbit circularization, 
may trigger AGN activity onto
the BHs,
at least temporary, and a stringent condition for
BH binary coalescence, in this context, may  come from the
request that the difference 
between the energy deposited by
the accreting  BHs in the gas and the energy
radiated away by cooling processes (here denoted as $\Delta E_{\rm acc}$)  
be less that the binding energy of
the gas itself (denoted with $E_{\rm iso,gas}$, here for simplicity): 
\begin{equation}
\Delta E_{\rm acc}\sim 
f_{\rm acc} \left ({L\over L_E}\right )
\left ({t\over t_{\rm Edd}}\right ) 
\epsilon M_{\rm BH}c^2<E_{\rm iso,gas}
\end{equation}
(with $L_E$  the Eddington luminosity and $f_{\rm acc}$ the fractional 
energy deposited by the accreting BH in the 
surrounding gas).
Typically  
\begin{equation}
\Delta E_{\rm acc}\approx 2\times 10^{59} 
f_{\rm rad} \left ({L\over L_E}\right )
\left ({t\over t_{\rm Edd}}\right )
\left ({\epsilon\over 0.1}\right ) 
\left ( {M_{\rm BH}\over 10^6\msun}\right ) \rm {erg}.
\end{equation}
\noindent 
This energy can be larger than $\Delta E_{\rm BBH},$ and may be comparable
to the binding energy of the environment, indicating that a major
threat to BH binary stalling in a gaseous background might come from
the radiative and/or mechanical energy emitted by the accreting BHs
during their inspiral and hardening.  Only short--lived (i.e., $t_{\rm
  GW}<t_{\rm Edd}$), and/or sub-Eddington accretion can guarantee the
persistence of a ``dense and cool'' gaseous structure around the
binary.  Modeling of realistic nuclear discs will help in quantifying
accurately $E_{\rm iso,gas}$ and the problem of the BH hardening in a
gaseous environment (Colpi et al. in preparation).

\section {Summary}

\noindent 
$\bullet$ In gas-rich galaxy--galaxy collisions, the BHs ``pair'' first
under the action of dynamical friction against the dark matter background.
When the merger is sufficiently advanced that tidal forces have
perturbed the gaseous/stellar discs of the mother galaxies, 
each BH is found to be surrounded by a prominent gaseous disc. 
The BHs have now separations of several kiloparsecs.

\noindent 
$\bullet$ When the merger is completed, a rotationally 
supported, turbulent, nuclear
gaseous disc forms at the center of the remnant galaxy.
In this ``grand'' disc of $\sim 19^9\,\msun,$
the BHs excite 
large scale density waves: gas--dynamical friction is the main driver
of their inspiral down to the parsec--scale.  
A few
million years after the completion of the merger
the BHs ``couple'' to form an eccentric {\it Keplerian binary}. The binary is
embedded in the typical, cool environment of a starburst.

\noindent 
$\bullet$ The binary eccentricity decreases to zero if the
BHs bind, in the grand disc, along a co--rotating orbit. It remains
large $(e>0)$, if the BHs bind along counter--rotating orbits.
However, this effect may not be generic: in
presence of a very massive nuclear disc 
with steep density profile, asymmetric instabilities growing in the
innermost self-gravitating region could in principle exert
torques on the binary, increasing its eccentricity.
The evolution of $e$ is thus sensitive to the dynamical
pattern of the gas surrounding the binary.  

\noindent 
$\bullet$ If the binary circularizes, a small-scale accretion disc
forms around each BH, and double AGN activity can be sustained for
approximately 1 Myr, on scales less than 10 parsecs.

\noindent 
$\bullet$ No compelling evidence exists on the actual fate of the BH
binary in a gaseous environment: whether it stalls or keep decaying
under the action of gravitational and viscous torques until
gravitational wave emission guides the inspiral toward coalescence.
This may depend sensitively on the BH mass, mass ratio, eccentricity,
and thermodynamical response of the gaseous environment to the BH
perturbation.  We note here that a possible threat may come from
energy injection by the BHs, should they accrete, or/and (to a lesser
extent) from the energy extracted from the orbit and deposited in the
surroundings.

\end{document}